\def\rfr#1{(\ref{#1})}
\def\rfrs#1#2{(\ref{#1})-(\ref{#2})}

\def\derp#1#2{\rp{\partial{#1}}{\partial{#2}}}
\def\dert#1#2{\frac{{{d}}{#1}}{{{d}}{#2}}}              

\def\asec{$''$ cy$^{-1}$}

\def\bar{\begin{eqnarray}}
\def\ear{\end{eqnarray}}
\def\bb{\bibitem}
\def\eqi{\begin{equation}}
\def\eqf{\end{equation}}
\def\eqia{\begin{eqnarray}}
\def\eqfa{\end{eqnarray}}
\def\rp#1#2{{#1\over#2}}
\def\ct#1{\cite{#1}}
\def\lb#1{\label{#1}}

\def\og{\omega}
\def\dfa{\derp{\mtc{R}}{a}}
\def\dfm{\derp{\mtc{R}}{\mtc{M}}}
\def\dfog{\derp{\mtc{R}}{\og}}
\def\dfi{\derp{\mtc{R}}{i}}
\def\dfe{\derp{\mtc{R}}{e}}
\def\dfo{\derp{\mtc{R}}{\Omega}}
\def\cu{\rp{2}{na}}
\def\cd{\rp{(1-e^2)}{na^{2} e}}
\def\ctr{\rp{\sqrt{1-e^2}}{na^{2} e}}
\def\cq{\rp{1}{na^{2}\sqrt{1-e^2}\si}}
\def\mtc#1{\mathcal{#1}}

\def\ci{\cos{i}}
\def\si{\sin{i}}

\documentclass[11pt]{article}
\usepackage{amsmath,amsthm,amscd,amssymb}
\usepackage{latexsym}
\usepackage{graphicx,epsfig}

\begin{document}

\noindent{\bf \LARGE{On the effects of the Dvali-Gabadadze-Porrati
braneworld gravity on the orbital motion of a test particle }}
\\
\\
\\
{\it Lorenzo Iorio\\
\\Viale Unit$\grave{a}$ di Italia 68, 70125\\Bari, Italy
\\e-mail: lorenzo.iorio@libero.it}

\begin{abstract}
In this paper we explicitly work out the secular perturbations
induced on all the Keplerian orbital elements of a test body to
order $\mathcal{O}(e^2)$ in the eccentricity $e$ by the weak-field
long-range modifications of the usual Newton-Einstein gravity due
to the Dvali-Gabadadze-Porrati (DGP) braneworld model. Both the
Gauss and the Lagrange perturbative schemes are used. It turns out
that the argument of pericentre $\omega$ and the mean anomaly
$\mathcal{M}$ are affected by secular rates which depend on the
orbital eccentricity via $\mathcal{O}(e^2)$ terms, but are
independent of the semimajor axis of the orbit of the test
particle. For circular orbits the Lue-Starkman (LS) effect on the
pericentre is obtained. Some observational consequences are
discussed for the Solar System planetary mean longitudes $\lambda$
which would undergo a $1.2\cdot 10^{-3}$ arcseconds per century
braneworld secular precession. According to recent data analysis
over 92 years for the EPM2004 ephemerides, the 1-sigma formal
accuracy in determining the Martian mean longitude amounts to
$3\cdot 10^{-3}$ milliarcseconds, while the braneworld effect over
the same time span would be $1.159$ milliarcseconds. The major
limiting factor is the $2.6\cdot 10^{-3}$ arcseconds per century
systematic error due to the mismodelling in the Keplerian mean
motion of Mars. A suitable linear combination of the mean
longitudes of Mars and Venus may overcome this problem. The
formal, 1-sigma obtainable observational accuracy would be $\sim
7\% $. The systematic error due to the present-day uncertainties
in the solar quadrupole mass moment $J_2$, the Keplerian mean
motions, the general relativistic Schwarzschild field and the
asteroid ring would amount to some tens of percent.
\end{abstract}
\section{Introduction}
Recently, a braneworld scenario which yields, among other things,
long-range modifications of the Newton-Einstein gravity has been
put forth by Dvali, Gabadadze and Porrati (DGP) \ct{DVP, Dvali
2004}. In such model, which encompasses an extra flat spatial
dimension, there is a free crossover parameter $r_0$, fixed by
observations to a value $\sim 5$ Gpc, beyond which traditional
gravity suffers strong modifications yielding to cosmological
consequences which would yield an alternative to the dark energy
in order to explain the observed acceleration of the Universe. The
consistency and stability of the DGP model have recently been
discussed in \ct{NicRat}.

Interestingly, for $R_{\rm g}<< r << r_0$, where $R_{\rm
g}=2GM/c^2$ is the usual Schwarzschild radius for a gravitating
body of mass $M$, there are also small modifications to the usual
Newton-Einstein gravity which could be detectable in a near
future. Indeed,  Lue and Starkman (LS) derived in \ct{Lue 2003} an
extra-pericentre advance for the orbital motion of a test particle
assumed to be nearly circular. Its magnitude is $\sim 4\cdot
10^{-4}$ arcseconds per century (\asec\ in the following). In
\ct{Lue 2003} it has been shown that the sign of such an effect is
related to the cosmological expansion phases allowed in this
model: the Friedmann-Lema\^{i}tre-Robertson-Walker (FLRW) phase
and the self-accelerating phase. The LS precession is an universal
feature of the orbital dynamics of a test particle because, in
this approximation, it is independent of its orbital parameters.

Since the present-day accuracy in measuring the non-Newtonian
perihelion rate of Mars amounts to $\sim 10^{-4}$ \asec\ (E.V.
Pitjeva, private communication 2004),  the possibility of
measuring such an effect in the Solar System scenario from the
planetary motion data analysis seems to be very appealing. It has
been investigated with some details in \ct{Iorio 2004} where a
suitable linear combination of the perihelia of some inner planets
has been considered.

In this paper we work out the secular effects of the DGP model on
all the Keplerian orbital elements of a test body to order
$\mathcal{O}(e^2)$ in the eccentricity $e$ with the Gauss and
Lagrange perturbative schemes. Possible observational implications
are worked out.

\section{The orbital effects}
\begin{figure}
\begin{center}
\includegraphics{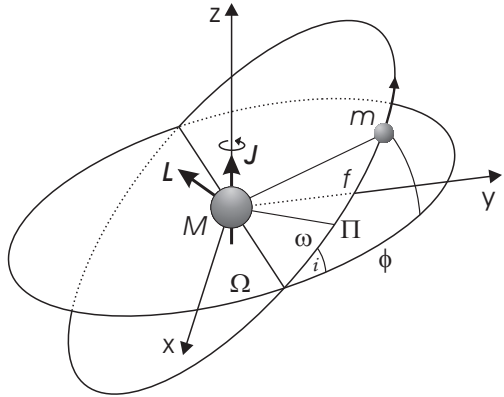}
\end{center}
\caption{\label{orbita}Orbital geometry for a motion around a
central mass. Here $L$ denotes the orbital angular momentum of the
particle of mass $m$, $J$ is the proper angular momentum of the
central mass $M$, $\Pi$ denotes the pericentre position, $f$ is
the true anomaly of $m$, which is counted from $\Pi$, $\Omega$,
$\omega$ and $i$ are the longitude of the ascending node, the
argument of pericentre and the inclination of the orbit with
respect to the inertial frame $\{x,y,z\}$ and the azimuthal angle
$\phi$ is the right ascension counted from the $x$ axis. When
orbits with small inclinations are considered, the longitude of
pericentre $\varpi=\Omega+\omega$ is used. }
\end{figure}

In this Section we will work out the secular effects of the DGP
gravity on the Keplerian orbital elements of the orbit of a test
body which is depicted in Figure \ref{orbita}.

From the metric for a static, spherical source in a cosmological
de Sitter background \ct{Lue 2003} \eqi (ds)^2=N^2(r,
w)(cdt)^2-A^2(r,w)(dr)^2-B^2(r,w)[(d\theta)^2+\sin^2\theta(d\phi)^2]-(dw)^2,\eqf
where $w$ is the fourth spatial coordinate,  and \eqi N\sim
1-\rp{R_{\rm g}}{2r}\pm\sqrt{\rp{rR_{\rm g}}{2r_0^2}},\eqf the
following Lagrangian can be obtained for a spherically symmetric
matter source located on the brane around the
origin\footnote{Since we are in the weak-field approximation, it
is assumed $N(r,w)=1+\tilde{n}(r,w),\ A(r,w)=1+\tilde{a}(r,w),\
B(r,w)=r[1+\tilde{b}(r,w)]$. The gauge $\tilde{b}(t, r)|_{w=0}=0$
has been adopted. The correction only to the Newtonian potential
has been considered, i.e. the spatial part of the metric has been
considered as Euclidean. Note that the modifications in the
cosmological background  introduced in \ct{newLS} do not alter
these results. } ($r=w=0$) \eqi \mathcal{L}_{\rm
DGP}=\rp{m}{2}\left[\left(1-\rp{R_{g}}{2r}\pm\sqrt{\rp{R_{\rm g
}r}{2r_0^2}}\right)^2 c^2-(\dot x )^2-(\dot y)^2-(\dot
z)^2\right].\lb{lagr}\eqf
\subsection{The Gauss perturbative scheme}\lb{Gauss}
From
\eqi\rp{d}{dt}\left({\derp{\mathcal{L}}{{\boldsymbol{v}}}}\right)-\derp{\mathcal{L}}{\boldsymbol{r}}=0\eqf
the resulting braneworld acceleration is
\eqi \boldsymbol{A}_{\rm DGP}=\mp
\left(\rp{c}{2r_0}\right)\sqrt{\rp{GM}{r}}\boldsymbol{\hat
r},\lb{dvali}\eqf where $\boldsymbol{\hat r}$ is the unit vector
in the radial direction. The term \rfr{dvali} can be regarded as a
small perturbation whose effects on the Keplerian orbital elements
of a test particle can be straightforwardly worked out, e.g., in
the Gauss perturbative scheme (e.g. \ct{roy}). Note that
\rfr{dvali} is purely radial.

The Gauss rate equations for the semimajor axis $a$, the
eccentricity $e$, the inclination $i$, the longitude of the
ascending node $\Omega$, the argument of pericentre $\omega$ and
the mean anomaly $\mathcal{M}$ of a test particle are
\bar
\dert{a}{t} & = & \rp{2}{n\sqrt{1-e^2}}\left[A_Re\sin f+A_T\left(\rp{p}{r}\right)\right],\lb{smax}\\
\dert{e}{t} & = & \rp{\sqrt{1-e^2}}{na}\left\{A_R\sin f+A_T\left[\cos f+\rp{1}{e}\left(1-\rp{r}{a}\right)\right]\right\},\\
\dert{i}{t} & = & \rp{1}{na\sqrt{1-e^2}}A_N\left(\rp{r}{a}\right)\cos (\omega+f),\lb{in}\\
\dert{\Omega}{t} & = & \rp{1}{na\sin i\sqrt{1-e^2}}A_N\left(\rp{r}{a}\right)\sin (\omega+f),\lb{nod}\\
\dert{\omega}{t} & = & -\cos i\dert{\Omega}{t}+\rp{\sqrt{1-e^2}}{nae}\left[-A_R\cos f+A_T\left(1+\rp{r}{p}\right)\sin f\right],\lb{perigeo}\\
\dert{\mathcal{M}}{t} & = & n
-\rp{2}{na}A_R\rp{r}{a}-\sqrt{1-e^2}\left(\dert{\omega}{t}+\cos
i\dert{\Omega}{t}\right),\lb{manom} \ear
in which $n\equiv 2\pi/P=\sqrt{GM/a^3}$ is the Keplerian mean
motion, $P$ is the test particle's orbital period, $f$ is the true
anomaly counted from the pericentre, $p=a(1-e^2)$ is the
semilactus rectum of the Keplerian ellipse, $A_R,\ A_T,\ A_N$ are
the in-plane radial, transverse and the out-of-plane components of
the perturbing acceleration, respectively, which have to be
evaluated on the unperturbed Keplerian ellipse  \eqi
r=\rp{a(1-e^2)}{1+e\cos f }.\lb{kep}\eqf The secular effects can
be obtained by averaging over one orbital period the
right-hand-sides of \rfrs{smax}{manom} evaluated on \rfr{kep}. The
average is performed by using
\eqi\rp{dt}{P}=\rp{(1-e^2)^{3/2}df}{2\pi(1+e\cos
f)^2}\lb{aver}\eqf and integrating from 0 to $2\pi$. Since the
perturbing acceleration is entirely radial, from \rfrs{in}{nod} it
is straightforward to note that there are no perturbations in the
inclination and the node. In regard to the other elements, the
following expansions have been performed \bar (1+e\cos f)^{-3/2}&
\sim & 1-\rp{3}{2}e\cos f+\rp{15}{8}e^2\cos^2
f-\rp{35}{16}e^3\cos^3
f,\lb{uno}\\
(1+e\cos f)^{-5/2}& \sim & 1-\rp{5}{2}e\cos f +\rp{35}{8}e^2\cos^2
f.\lb{due} \ear The expansion \rfr{uno} has been used for the
semimajor axis, the eccentricity and the pericentre, while the
expansion \rfr{due} has been used for the mean anomaly.

It turns out that only the pericentre and the mean anomaly undergo
secular effects which are, to order $\mathcal{O}(e^2)$ \bar
\dert{\omega}{t}&\sim &\mp\rp{3c}{8
r_0}\left(1-\rp{13}{32}e^2\right),\lb{LS}\\
\dert{\mathcal{M}}{t}& \sim
&\pm\rp{11c}{8r_0}\left(1-\rp{39}{352}e^2\right).\lb{gaussmanom}\ear
For $e\rightarrow 0$ \rfr{LS} reduces to the result by Lue and
Starkman
 for circular orbits \ct{Lue 2003}.
\subsection{The Lagrangian perturbative
scheme}\lb{Lagra} The Lagrange planetary equations for the rates
of change of the Keplerian orbital elements are (e.g. \ct{Kaula
1966})\bar
\dert{a}{t}&=& \cu \ \dfm ,\lb{lag_sma}\\
\dert{e}{t}&=& \cd \ \dfm - \ctr \ \dfog ,\lb{lag_ecce}\\
\dert{i}{t}&=& \ci \cq \ \dfog - \cq \ \dfo ,\lb{lag_incli}\\
\dert{\Omega}{t}&=& \cq \ \dfi ,\lb{lag_omeg}\\
\dert{\og}{t}&=&- \ci \cq \ \dfi + \ctr \ \dfe ,\lb{lag_nod}\\
\dert{\mtc{M}}{t}&=&n- \cd \ \dfe - \cu \ \dfa ,\lb{lag_manom}
 \ear
 where $\mathcal{R}$ is the perturbing function which accounts for
all the departures of the gravitational potential from the
Newtonian monopole term. Such a perturbing scheme was applied for
the first time  to post-Newtonian motions by Rubincam
\ct{Rubincam}. In the case of the DGP braneworld theory from
$\mathcal{H}=\boldsymbol{p}\cdot\boldsymbol{v}-\mathcal{L}$,
$\boldsymbol{p}=\partial\mathcal{L}/\partial\boldsymbol{v}$, and
\rfr{lagr} it can be obtained \eqi\mathcal{R}_{\rm DGP}=
\mp\rp{c}{r_0}\sqrt{GMr}.\lb{ham} \eqf  The DGP perturbing
function \rfr{ham} has to be evaluated on the unperturbed
Keplerian ellipse \rfr{kep} and averaged over one orbital
revolution by means of \rfr{aver} in order to obtain the secular
effects. From \eqi(1+e\cos f)^{-5/2}\sim 1-\rp{5}{2}e\cos
f+\rp{35}{8}e^2\cos^2 f-\rp{105}{16}e^3\cos^3
f+\rp{1155}{128}e^4\cos^4 f,\eqf one obtains
\eqi\left\langle\mathcal{R}\right\rangle\sim
\mp\rp{c\sqrt{GMa}}{r_0}\left(1+\rp{3}{16}e^2+
\rp{9}{1024}e^4\right).\eqf Since
\eqi\derp{\left\langle\mathcal{R}\right\rangle}{\mathcal{M}}=
\derp{\left\langle\mathcal{R}\right\rangle}{\Omega}=\derp{\left\langle\mathcal{R}\right\rangle}{\omega}=
\derp{\left\langle\mathcal{R}\right\rangle}{i}=0, \eqf is it
straightforward to obtain
\eqi\dert{a}{t}=\dert{e}{t}=\dert{i}{t}=\dert{\Omega}{t}=0.\eqf
The situation is different for the pericentre and the mean
anomaly. Indeed,
\bar\derp{\left\langle\mathcal{R}\right\rangle}{e}&=&
\mp\rp{3ec\sqrt{GMa}}{8r_0}\left(1+\rp{3}{32}e^2\right),\\
\derp{\left\langle\mathcal{R}\right\rangle}{a}&=&\mp\rp{c}{2r_0}\sqrt{\rp{GM}{a}}\left(1+\rp{3}{16}e^2+
\rp{9}{1024}e^4\right) .\ear As a consequence,
\bar\dert{\omega}{t}&\sim&\mp\rp{3c}{8
r_0}\left(1-\rp{13}{32}e^2\right),\lb{lagraperi}\\
\dert{\mathcal{M}}{t}&\sim&\pm\rp{11c}{8
r_0}\left(1-\rp{39}{352}e^2\right).\lb{lagramanom}\ear Note that
\rfrs{lagraperi}{lagramanom} are identical to
\rfrs{LS}{gaussmanom} obtained in Section \ref{Gauss}.
\section{The possible use of the planetary mean longitudes}
In \ct{Iorio 2004} only the use of the planetary perihelia has
been examined. First, preliminary observational tests are
discussed in \ct{Iorioau}.

The fact that also the mean anomaly is affected by the DGP gravity
with a relatively large effect suggests to examine the possibility
of using the planetary mean longitudes
$\lambda=\omega+\Omega+\mathcal{M}$.
\subsection{The observational sensitivity}
The sizes of the DGP secular precessions and of many Newtonian and
Einsteinian competing secular precessions of the mean longitudes
of the Solar System planets are reported in Table \ref{sensi}.
\begin{table}
\caption{Nominal values, in \asec, of the secular precessions
induced on the planetary mean longitudes $\lambda$ by the DGP
gravity and by some of the competing Newtonian and Einsteinian
gravitational perturbations. For a given planet, the precession
labelled with Numerical includes all the numerically integrated
perturbing effects of the dynamical force models used at JPL for
the DE200 ephemerides. E.g., it also comprises the classical N-
body interactions, including  the Keplerian mean motion $n$. For
the numerically integrated planetary precessions see
http://ssd.jpl.nasa.gov/elem$\_$planets.html$\#$rates. The effect
labelled with GE is due to the post-Newtonian general relativistic
gravitoelectric Schwarzschild component of the solar gravitational
field, that labelled with $J_2$ is due to the classical effect of
the Sun's quadrupole mass moment $J_2$ and that labelled with LT
is due to the post-Newtonian general relativistic gravitomagnetic
Lense-Thirring \ct{Lense and Thirring 1918} component of the solar
gravitational field (not included in the force models adopted by
JPL). For $J_2$ the value $1.9\cdot 10^{-7}$ has been adopted
\ct{Pitjeva 2004}. For the Sun's proper angular momentum $J$,
which is the source of the gravitomagnetic field, the value
$1.9\cdot 10^{41}$ kg m$^2$ s$^{-1}$ \ct{Pijpers 2003} has been
adopted. }\label{sensi}
\begin{tabular}{@{\hspace{0pt}}llllll}
\hline\noalign{\smallskip} Planet & DGP & Numerical & GE & $J_2$ &
LT
\\
\noalign{\smallskip}\hline\noalign{\smallskip} Mercury & $1.2\cdot
10^{-3}$ & $5.381016282\cdot 10^8$ & $-8.48\cdot 10^1$ & $4.7\cdot
10^{-2}$ & $-2\cdot
10^{-3}$\\
Venus & $1.2\cdot 10^{-3}$ & $2.106641360\cdot 10^8$& $-1.72\cdot
10^1$ &$5\cdot 10^{-3}$ & $-3\cdot 10^{-4}$\\
Earth & $1.2\cdot 10^{-3}$ & $1.295977406\cdot 10^8$ & $-7.6$ &
$1.6\cdot 10^{-3}$& $-1\cdot 10^{-4}$\\
Mars & $1.2\cdot 10^{-3}$ & $6.89051037\cdot 10^7$ & $-2.6$
&$3\cdot
10^{-4}$& $-3\cdot 10^{-5}$\\
Jupiter & $1.2\cdot 10^{-3}$ & $1.09250783\cdot 10^7$ & $-1\cdot
10^{-1}$ & $5\cdot 10^{-6}$ & $-7\cdot 10^{-7}$\\
Saturn & $1.2\cdot 10^{-3}$ & $4.4010529\cdot 10^6$ & $-2\cdot
10^{-2}$ & $6\cdot 10^{-7}$ & $-1\cdot 10^{-7}$\\
Uranus & $1.2\cdot 10^{-3}$ & $1.5425477\cdot 10^{6}$ &$-4\cdot
10^{-3}$ & $5\cdot 10^{-8}$ & $-1\cdot 10^{-8}$\\
Neptune & $1.2\cdot 10^{-3}$ & $7.864492\cdot 10^{5}$ & $-1\cdot
10^{-3}$ & $1\cdot 10^{-8}$ & $-5\cdot 10^{-9}$\\
 \noalign{\smallskip}\hline
\end{tabular}
\end{table}
It can be noted that the DGP secular precessions amount to
$1.2\cdot 10^{-3}$ \asec\ for all the planets: the corrections due
to the eccentricities are very small amounting to almost $10^{-5}$
\asec. In regard to the observational sensitivity, in Table
\ref{accu}, retrieved from Table 4 of \ct{Pitjeva 2004}, the most
recent results for the EPM2004 ephemerides are presented.
\begin{table}
\caption{Formal standard deviations, in milliarcseconds (mas), of
the planetary mean longitudes as from Table 4 of \ct{Pitjeva 2004}
for the EPM2004 ephemerides. Note that realistic errors may be an
order of magnitude larger. About 300000 position observations
(1911-2003) of different types (optical, radar, spacecraft, LLR)
have been used. The braneworld shift for $\lambda$ amounts to
1.159 mas over a 92-years time span. }\label{accu}
\begin{tabular}{@{\hspace{0pt}}llllllll}
\hline\noalign{\smallskip} Mercury & Venus & Mars & Jupiter &
Saturn & Uranus & Neptune
\\
\noalign{\smallskip}\hline\noalign{\smallskip} $3.75\cdot 10^{-1}$
& $1.87\cdot 10^{-1}$ & $3\cdot 10^{-3}$ & 1.109 & 3.474 & 8.818 &
$3.5163\cdot 10^1$\\
 \noalign{\smallskip}\hline
\end{tabular}
\end{table}
They are based on the processing of a vast amount of data of
different kinds (optical, radar, spacecraft, LLR) ranging from
1911 to 2003. It can be noted that Mars is the best candidate for
extracting the DGP effect because the formal standard deviation in
$\lambda_{\rm Mars}$ amounts to $3\cdot 10^{-3}$ milliarcseconds
(mas) only while the braneworld shift for the same time span is
1.159 mas.
\section{Some
systematic errors} In this Section we will examine various
competing classical and general relativistic effects which would
act as sources of systematic errors in order to see if it is
possible to use only the Martian mean longitude for the proposed
test.
\subsection{The impact of the solar quadrupole
mass moment and of the Einstein gravitoelectric force} In regard
to the systematic errors which would induced by the other
Newtonian and Einsteinian competing effects, the solar quadrupole
mass moment $J_2$ is presently known with a 15$\%$ accuracy (Table
7 of \ct{Pitjeva 2004}), so that the mismodelled part of its
secular precession would be two orders of magnitude smaller than
the effect of interest.

More important would be the impact of the Einsteinian
gravitoelectric precession. Indeed, the general relativistic
perihelion precession has been measured to a $10^{-4}$ relative
accuracy via the PPN parameters $\beta$ and $\gamma$ (Table 8 of
\ct{Pitjeva 2004}): assuming that it would also hold for
$\lambda$, this would yield a $2\cdot 10^{-4}$ \asec\ mismodelled
effect. However, it should be noted that these are merely the
formal 1-sigma errors: realistic bounds might be one order of
magnitude larger.
\subsection{The N-body
perturbations} The mean longitude can be written as
$\lambda=nt+\epsilon,$ so that \ct{muder99}
\eqi\dert{\lambda}{t}=n+\dert{\epsilon^{\ast}}{t}\equiv
n+\dert{n}{t}t+\dert{\epsilon}{t},\eqf where
\eqi\dert{\epsilon}{t}=-\rp{2}{na}\derp{\mathcal{R}}{a}+\rp{\sqrt{1-e^2}(1-\sqrt{1-e^2})}{na^2e}
\derp{\mathcal{R}}{e}+\rp{\tan(i/2)}{na^2\sqrt{1-e^2}}\derp{\mathcal{R}}{i}.\eqf
\subsubsection{The direct and indirect effects on
the Keplerian mean motion due to the other planets and the
asteroids} In regard to the Keplerian mean motion, it turns out to
be the major limiting factor. Indeed, the uncertainty in the solar
$GM$ $\sigma_{GM_{\odot}}=8\cdot 10^9$ m$^3$ s$^{-2}$ (Table 1 of
\ct{Standish 1995})  yields $\sigma_{n_{\rm Mars}}=2.6\cdot
10^{-3}$ \asec. However, if we adopt the position of keeping fixed
the solar $GM$, the formal 1-sigma error due to the semimajor axis
$\sigma_{a_{\rm Mars}}=6.57\cdot 10^{-1}$ m (Table 4 of
\ct{Pitjeva 2004}) only reduces to $\sigma_{n_{\rm Mars}}=3\cdot
10^{-4}$ \asec.

Another source of potential bias is represented by the indirect
perturbations on the Keplerian mean motion induced by the
variations in the semimajor axis \eqi\Delta
n\equiv\dert{n}{t}t=-\rp{3}{2}\rp{n}{a}\dert{a}{t}t=-\rp{3}{a^2}\derp{\langle\mathcal{R}\rangle}{\lambda}t.
\eqf
According to \ct{muder99}, there are no secular effects on the
semimajor axis: instead, the so called resonant perturbations,
affect this Keplerian orbital element. They are induced by those
terms in the expansion of the N-body disturbing function which
retain the mean longitudes of the perturbed and the perturbing
bodies. Such kind of harmonic perturbations, due to the asteroids
for Mars, are potentially very insidious because they may have
large amplitudes and extremely long periods.  This topic has been
treated in \ct{jim}. Table IV of \ct{jim} lists the most important
of such perturbations. The nominal amplitudes of the perturbations
induced, e.g., by (1) Ceres, (2) Pallas, (4) Vesta and (7) Iris
are $4.7\cdot 10^{-2}$ \asec, $1.2\cdot 10^{-2}$ \asec, $5.7\cdot
10^{-2}$ \asec\ and $5\cdot 10^{-3}$ \asec, respectively.
According to the results of Table 6 of \ct{Pitjeva 2004}, their
mismodelled parts amount to, $7\cdot 10^{-5}$ \asec, $3\cdot
10^{-5}$ \asec, $4\cdot 10^{-5}$ \asec\ and $8\cdot 10^{-5} $
\asec, respectively. It must also be noted that the integrated
shift of $\Delta n$ grows quadratically in time.
\subsubsection{The perturbations on
the mean longitude due to the other planets and the asteroid ring}
Here we will deal with $d\epsilon/dt$ which is responsible for the
secular perturbations on $\lambda$.

The Newtonian secular perturbations induced on the Mars mean
longitude by the other planets of the Solar System are of the
order of $10^3$ \asec. The major source of uncertainty is
represented by the $GM$ of the perturbing bodies among which
Jupiter plays a dominant role, especially for Mars. According to
\ct{Jacobson 2003}, the Jovian $GM$ is known with a relative
accuracy of $10^{-8}$; this would imply for the red planet a
mismodelled precession induced by Jupiter of the order of
$10^{-5}$ \asec. The $GM$ of Saturn is known with a $1\cdot
10^{-6}$ relative accuracy \ct{Jacobson 2004}. However, the ratio
of the secular precession induced on the Martian Keplerian
elements by Saturn to that induced on the Mars perihelion by
Jupiter is proportional to $(M_{\rm Sat}/M_{\rm Jup})(a_{\rm
Jup}/a_{\rm Sat })^3\sim 5\cdot 10^{-2}$. This would assure that
also the effect of Saturn is of the order of $10^{-5}$ \asec. The
situation with the precessions induced by Uranus and Neptune is
even more favorable. Indeed, for Uranus $(M_{\rm Ura}/M_{\rm
Jup})(a_{\rm Jup}/a_{\rm Ura })^3\sim 1\cdot 10^{-3}$ and the
relative uncertainty in the uranian $GM$ is $2\cdot 10^{-6}$
\ct{Jacobson 1992}. For Neptune $(M_{\rm Nep}/M_{\rm Jup})(a_{\rm
Jup}/a_{\rm Nep })^\sim 3\cdot 10^{-4}$ and $\sigma_{GM}/GM=2\cdot
10^{-6}$ \ct{Jacobson 1991}.

A source on potentially non-negligible perturbations on the
Martian mean longitude is the asteroid ring, i.e. the ensemble of
the minor asteroids whose impact can be modelled as due to a solid
ring in the ecliptic plane \ct{kras}. The perturbations due to it
can be worked out, e.g., with the Lagrangian approach and the
disturbing function of the Appendix of \ct{kras}. By using the
values of \ct{Pitjeva 2004} for the ring's radius and mass it
turns out that the secular perturbation on $\dot\lambda_{\rm
Mars}$ amounts to $-3.4\cdot 10^{-3}$ \asec, with an uncertainty
of $3\cdot 10^{-4}$ \asec.
\subsection{The total systematic error on the mean longitude of Mars}
In Table \ref{syserr} we summarize the various systematic errors
affecting the mean longitude of Mars.
\begin{table}
\caption{Sources of systematic errors, in \asec, affecting the
mean longitude of Mars. The reported figures are at 1-sigma level.
The total effect, obtained by summing up the various errors, is
more than 90$\%$ of the braneworld signature. }\label{syserr}
\begin{tabular}{@{\hspace{0pt}}ll}
\hline\noalign{\smallskip} Source of systematic error &
Mismodelled amplitude (\asec)
\\
\noalign{\smallskip}\hline\noalign{\smallskip} Keplerian mean
motion & $3\cdot 10^{-4}$\\
Asteroid ring & $3\cdot 10^{-4}$\\
Schwarzschild GE field & $2\cdot 10^{-4}$\\
(7) Iris & $8\cdot
10^{-5}$\\
(1) Ceres & $7\cdot
10^{-5}$\\
Solar $J_2$ & $4.5\cdot 10^{-5}$\\
(4) Vesta & $4\cdot
10^{-5}$\\
(2) Pallas & $3\cdot
10^{-5}$\\
Lense-Thirring GM field (assumed unmodelled) & $3\cdot 10^{-5}$\\
\noalign{\smallskip}\hline\noalign{\smallskip} Total & $1.1\cdot 10^{-3}$\\
 \noalign{\smallskip}\hline
\end{tabular}
\end{table}
It turns out that it is impossible to only analyze the mean
longitude of Mars: indeed, the 1-sigma total error would amounts
to $\sim 1.1\cdot 10^{-3}$ \asec, i.e. more than 90$\%$ of the
braneworld effect.

In the next Section we will outline a possible strategy to
suitably combine the data of Mars with those of other inner
planets in order to cancel out, by construction, many systematical
error. A numerical example is explicitly worked out.
\section{The linear combination approach}
In order to cancel out the impact of the various sources of
systematic errors it is possible to suitably linearly combine the
mean longitudes of Mars and Venus following an approach adopted
in, e.g., \ct{2005a, 2005b}. Let us assume we have at our disposal
$N$ Keplerian orbital elements\footnote{They could all belong to
the same planet or, alternatively, they could all be the same
element, say, the mean longitude, of $N$ planets or a mix of these
possibilities. } $\mathcal{K}$ whose time evolution is supposed to
be affected by a certain number of Newtonian and post-Newtonian
effects, say \eqi\dot{\mathcal{K}} =\dot{\mathcal{K}}_{\rm DGP
}+\dot{\mathcal{K}}_{J_{2}}+\dot{\mathcal{K}}_{\rm GE
}+\dot{\mathcal{K}}_{\rm N-body}+\dot{\mathcal{K}}_{\rm
LT}+...\lb{kappa}.\eqf If we are interested in isolating one
particular feature, say $\dot{\lambda}_{\rm DGP}$, and we know it
is smaller than other larger effects which affect the same
Keplerian element we can explicitly write down the expressions of
the observational residuals\footnote{Here we speak about residuals
of Keplerian orbital elements in a, strictly speaking, improper
sense. The Keplerian orbital elements are not directly observable:
they can only be computed. The basic observable quantities are
ranges, range-rates and angles. Here we mean the differences
between the time series of $\mathcal{K}$ obtained from a given
observed orbital arc and the time series of $\mathcal{K}$ obtained
from a propagated orbital arc with the force we want to cancel in
the force models. The two time series share the same (measured)
initial conditions.} $\delta\dot{\mathcal{K}}_{\rm obs}$ in term
of the feature of interest- which will be assumed to be entirely
(or partly) present in the residuals-and of the main larger
aliasing effects-which will affect the residuals with their
mismodelled part only-so that the number of terms in the sum in
the right-hand-side of \rfr{kappa} which represent the effect of
interest and the other most relevant larger bias is equal to the
number $N$ of Keplerian orbital elements we have at our disposal.
Now we have a system of $N$ equations in $N$ unknowns which we can
solve for the effect we are interested in. The resulting
expression will be, by construction, independent of the other
larger aliasing effects.

By using, e.g., the figures of Table \ref{sensi} for the
numerically integrated total precessions, which encompasses all
the Newtonian and general relativistic effects, it is possible to
obtain \eqi \delta\dot\lambda_{\rm Mars}+c_1\delta\dot\lambda_{\rm
Venus}= \mp 8\cdot 10^{-4}\ ''{\rm cy}^{-1},\lb{combinaz}\eqf
where \eqi c_1=-\rp{\dot\lambda^{(\rm Num)}_{\rm
Venus}}{\dot\lambda^{(\rm Num )}_{\rm Mars}}=-3.270\cdot 10^{-1}
,\eqf and $\delta\dot\lambda$ are the time series residuals of the
mean longitude built up in  order to entirely absorb all the
non-Newtonian and non-Einsteinian gravity.

The combination \rfr{combinaz} is not affected, by construction,
by all the competing Newtonian and general relativistic
perturbations acting upon $\lambda$, at least to the level of
accuracy of the dynamical force models used in calculating
$\dot\lambda^{(\rm Num)}$. Of course, the coefficient $c_1$ can,
in principle, be more precisely recalculated  by using future,
more accurate ephemerides. Indeed, from the previous discussion
should be clear the importance of also including, at the best of
our knowledge, the asteroids. While in the mathematical model of
DE200 the perturbations of only three of the major asteroids are
present, in the more advanced DE410 and EPM2004 ephemerides the
perturbations of 300 asteroids and also of the asteroid ring have
been included.

From Table \ref{accu} it is possible to obtain for the formal
1-sigma observational accuracy in \rfr{combinaz} a 7$\%$ value
over 92 years. The systematic error due to the solar quadrupole
mass moment, the Keplerian mean motions, the general relativistic
Schwarzschild field and the asteroid ring amounts to some tens
percent by assuming the formal, 1-sigma level of uncertainty of
\ct{Pitjeva 2004} for $J_2$, $a_{\rm Mars}$, $a_{\rm Venus}$,
$\beta$, $\gamma$ and $M_{\rm ring}$.
\section{The possibility of using the proposed LARES/WEBER-SAT satellite}

The Earth artificial satellite LARES/WEBER-SAT \ct{Ciufolini 1986,
Iorio et al. 2002} was proposed in order to measure the
Lense-Thirring effect
 on the orbit of a test particle to a
high accuracy level ($\sim 1\%$) in the gravitational field of the
Earth by suitably combining its data with those of the existing
geodetic SLR (Satellite Laser Ranging) satellites LAGEOS and
LAGEOS II. Their orbital parameters are in Table \ref{para}.
\begin{table}
\caption{Orbital parameters of LAGEOS, LAGEOS II and LARES.
}\label{para}
\begin{tabular}{@{\hspace{0pt}}llll}
\hline\noalign{\smallskip}
 Orbital element & LAGEOS & LAGEOS II & LARES
\\
\noalign{\smallskip}\hline\noalign{\smallskip}
semimajor axis $a$ (km) & $1.2270\cdot 10^4$ & $1.2163\cdot 10^4$ & $1.2270\cdot 10^4$\\
eccentricity $e$ & $4.5\cdot 10^{-3}$ & $1.4\cdot 10^{-2}$ & $4.0\cdot 10^{-2}$ \\
inclination $i$ (deg) & $1.10\cdot 10^2$ & $5.265\cdot 10^1$ & $7.0\cdot 10^1$ \\
\noalign{\smallskip}\hline
\end{tabular}
\end{table}
In \ct{Ciufolini 2004} Ciufolini has proposed to measure the
braneworld effect on the pericentre, which amounts to $\sim 4\cdot
10^{-3}$ milliarcseconds per year (mas yr$^{-1}$), with the
perigee of LARES. Unfortunately, it would be impossible, as we
will show in the following.
\subsection{The observational sensitivity} From $\Delta
r\sim ea\Delta\omega$ \ct{Nordtvedt 2000}, it can be obtained that
the accuracy in measuring the perigee precession over a given
observational time span can be expressed as
$\delta\omega\sim\delta r/ea$. By assuming a root-mean-square
(rms) error of 1 mm in reconstructing the LARES orbit over, say,
one year for a given set of dynamical force models one gets
$\delta\omega=4\cdot 10^{-1}$ mas.

With\footnote{It may be interesting to note that, with such an
orbital configuration and $i=6.34\cdot 10^1$ deg the
gravitomagnetic Lense-Thirring precession would amount to -1.8 mas
yr$^{-1}$ only.} $a=3.6\cdot 10^4$ km and $e=2.8\cdot 10^{-1}$, as
also suggested in \ct{Ciufolini 2004}, and by assuming a rms error
of 1 cm, the accuracy in the perigee would amount to $2\cdot
10^{-1}$ mas. Note that the adopted values for the obtainable
accuracies in $r$ are optimistic; for example, the mm accuracy has
not yet been fully achieved for the existing LAGEOS satellites.

\subsection{The
systematic errors of gravitational origin} The perigee of an Earth
artificial satellite is affected by various kinds of long-period
(i.e. averaged over one orbital revolution) orbital perturbations
induced by the multipolar expansion of the Earth's gravitational
potential \ct{Kaula 1966}. The most insidious ones are the secular
rates induced by the even ($\ell=2,4,6...$) zonal ($m=0$) harmonic
coefficients $J_{\ell}$ of the geopotential which account for the
departure of the Earth from an exact spherical shape. Their
explicit expressions up to degree $\ell=20$ can be found, e.g., in
\ct{Iorio 2003}. The largest precession is induced by the Earth's
quadrupole mass moment $J_2$. For a moderate eccentricity its
analytical expression is
\eqi\left.\frac{d\omega}{dt}\right|_{J_{2}}=\rp{3}{2}n\left(\frac{R}{a}\right)^2\rp{J_2}{(1-e^2)^2}
\left(2-\frac{5}{2}\sin^2 i \right),\lb{prece}\eqf where $R$ is
the Earth's equatorial mean radius.

In \ct{Ciufolini 2004} it is proposed to launch LARES in the so
called frozen-perigee orbit characterized by the critical value of
the inclination, $i=6.34\cdot 10^1$ deg, for which the
$J_2$-precession of \rfr{prece} vanishes. Moreover, it seems that
Ciufolini suggests to use the LARES data together with those from
LAGEOS and LAGEOS II in order to measure the Lense-Thirring as
well.

We will now show that such proposals are unfeasible.
\subsubsection{The impact of the even zonal harmonics of the
geopotential} Indeed, apart from the fact that the unavoidable
orbital injection errors would prevent to exactly insert LARES in
orbit with the required inclination, it turns out that the impact
of the other uncancelled precessions induced by the even zonal
harmonics  of higher degree, along with their secular variations,
would totally swamp the LS effect. In Table \ref{orpar} we use the
calibrated sigmas of the even zonal harmonics of  the recently
released combined CHAMP+GRACE+terrestrial gravity EIGEN-CG01C
Earth gravity model \ct{Reigber et al. 2004} in order to calculate
the mismodelled residual classical precessions on the perigee of
LARES by assuming $i=6.34\cdot 10^1$ deg.
\begin{table}
\caption{Mismodelled classical secular precessions, in mas
yr$^{-1}$, of the perigee of an Earth satellite with semimajor
axis $a=1.2270\cdot 10^4$ km, inclination $i=6.34\cdot 10^1$ deg,
eccentricity $e=4.0\cdot 10^{-2}$ according to the combined
CHAMP+GRACE+terrestrial gravity EIGEN-CG01C Earth gravity model up
to degree $\ell=20$ (1-sigma). For the precessions induced by the
secular variations of the even zonal harmonics, referred to an
observational time span of one year, the values for $\dot
J_{\ell}, \ell=2,4,6$ of \ct{Cox et al. 2003} have been used. Such
precessions grow linearly in time.  In the last three rows a - has
been inserted because the corresponding mismodelled precessions
amount to $10^{-4}$ mas yr$^{-1}$. The LS secular precession
amounts to $\mp 4\cdot 10^{-3}$ mas yr$^{-1}$ for an object
orbiting a central mass along a circular path. }\label{orpar}

\vspace{0.5cm}
\begin{tabular}{@{\hspace{0pt}}ll}
\hline\noalign{\smallskip} Even zonal harmonic & Mismodelled
precessions (mas yr$^{-1}$)
\\
\noalign{\smallskip}\hline\noalign{\smallskip}
$J_2$ & $1.26\cdot 10^{-1}$\\
$\dot J_2$ & $4\cdot 10^{-3}$ mas yr$^{-2}$\\
$J_4$ & 4.390\\
$\dot J_4$ & 1.406 mas yr$^{-2}$\\
$J_6$ & 1.355\\
$\dot J_6$ & $6.63\cdot 10^{-1}$ mas yr$^{-2}$\\
$J_8$ & $2.08\cdot 10^{-1}$\\
$J_{10}$ & $2.5\cdot 10^{-2}$\\
$J_{12}$ & $2.5\cdot 10^{-2}$\\
$J_{14}$ & $7\cdot 10^{-3}$\\
$J_{16}$ & -\\
$J_{18}$ & -\\
$J_{20}$ & -\\

\noalign{\smallskip}\hline
\end{tabular}
\end{table}
It can be easily seen that the mismodelled precessions induced by
the first seven even zonal harmonics are larger than the LS
precession. Due to the extreme smallness of such an effect, it is
really unlikely that the forthcoming Earth gravity models from
CHAMP and GRACE will substantially change the situation. This
rules out the possibility of using only the perigee of LARES.

\subsubsection{The linear combination approach} In regard to the
possibility of suitably combining the Keplerian orbital elements
of the existing LAGEOS satellites and of the proposed LARES
\ct{Ciufolini 1996, Iorio 2005, Iorio and Doornbos 2004} in order
to reduce the impact of the systematic errors of the geopotential
on the proposed measurement, it is unfeasible as well.

Indeed, the perigee is also affected, among other things, by the
 Einstein precession
\ct{Einstein 1915}, whose nominal value for LARES is
$3.280136\cdot 10^3$ mas yr$^{-1}$, and by the Lense-Thirring
effect, which, for $i=6.34\cdot 10^1$ deg, nominally amounts to
-$4.1466\cdot 10^1$ mas yr$^{-1}$. This means that if we want to
measure the LS precession independently of such quite larger
Newtonian and post-Newtonian effects we would need ten Keplerian
orbital elements in order to write down a linear system of ten
equations in ten unknowns (the first seven even zonal harmonics,
the LS effect and the two relativistic precessions) and solve it
for the LS precession. Instead, we would have at our disposal, in
principle, only four reliable Keplerian orbital
elements\footnote{Note that the other routinely and accurately
laser-tracked  SLR satellites which could, in principle, be
considered are Ajsai, Starlette and Stella: the useful orbital
elements are their nodes and the perigee of Starlette. However,
since they orbit at much lower altitudes than the LAGEOS
satellites they would practically be useless. Indeed, they are
sensitive to much more even zonal harmonics of the geopotential so
that they would greatly increase the systematic error induced by
them.}: the  nodes $\Omega$ of LAGEOS, LAGEOS II and LARES-which
are affected by the Earth's geopotential and by the Lense-Thirring
effect but not by the LS force-and the perigee of LARES. They
would only allow to cancel out the general relativistic effects
and just one even zonal harmonic.

The data from an hypothetical satellite with $a=3.6\cdot 10^4$ km,
as proposed in \ct{Ciufolini 2004}, could not be used for the
following reasons. If, on the one hand, the classical geopotential
precessions would be smaller than the LS effect, apart from those
induced by $J_4$ and $J_6$ according to EIGEN-CG01C, on the other
hand, the gravitomagnetic Lense-Thirring and the gravitoelectric
Einstein post-Newtonian precessions on the perigee would amount to
-1.8512 mas yr$^{-1}$ (for $i=6.34\cdot 10^1$ deg) and
$2.409956\cdot 10^2$ mas yr$^{-1}$, respectively. This means that,
even for such a higher altitude, one could not analyze only the
perigee of LARES whose Keplerian orbital elements should be,
instead, combined with those of the existing LAGEOS and LAGEOS II:
with the nodes of the LARES, LAGEOS and LAGEOS II and the perigee
of LARES it would be possible, in principle, to disentangle the LS
effect from the post-Newtonian precessions and one classical even
zonal harmonic. For such a high altitude the period of the node of
LARES would amount to $\sim 10^4$ days, i.e. tens of years. The
tesseral $K_1$ tidal perturbation, which is one of the most
powerful harmonic time-dependent perturbations which are not
cancelled out by the linear combination approach, has just the
period of the node. Then, it would act as a superimposed linear
bias over an observational time span of a few years. Moreover, as
shown in \ct{Iorio 2002, Vespe and Rutigliano 2004}, when
high-altitude satellites are included in linear combinations
involving also lower satellites as the existing LAGEOS and LAGEOS
II it turns out that the orbital elements of the higher SLR
targets enter the combination with huge coefficients which amplify
all the uncancelled orbital perturbations. This would also be the
case for the $K_1$ tide affecting LARES.
\subsubsection{The impact of the odd zonal
harmonics} The perigee of an Earth artificial satellite is also
affected by long-period harmonic perturbations induced by the odd
($\ell=3,5,7...$) zonal ($m=0$) harmonics of the geopotential. The
largest perturbation is induced by $J_3$: it has a sinusoidal
signature with the period of the perigee. Its analytic expression
is \ct{J3}
\begin{eqnarray}
\left.\rp{d\omega}{dt}\right|_{J_{3}}&=&-\rp{3}{2}n\left(\frac{R}{a}\right)^3\rp{J_3\sin\omega}{e\sin
i (1-e^2)^3}\left[\left(\rp{5}{4}\sin^2 i-1\right)\sin^2
i+\right.\nonumber\\
&+&\left.e^2\left(1-\rp{35}{4}\sin^2 i\cos^2
i\right)\right].\lb{j3}
\end{eqnarray}
 For $i=6.34\cdot 10^1$ deg the period
of the perigee of LARES, given by \rfr{prece}, is of the order of
$10^5$ days; moreover, the second term of the right-hand-side of
\rfr{j3} does not vanish. This means that the perigee of LARES in
the critical inclination would be affected by an additional
semisecular bias due to $J_3$ which, over an observational time
span of some years, would resemble a superimposed linear trend.
According to EIGEN-CG01C, its mismodelled effect would be $\leq
5\cdot 10^{-1}$ mas yr$^{-1}$.

This additional bias should be accounted for both in the
perigee-only scenario and in the linear combinations scenario.
Note also that for $a=3.6\cdot 10^4$ km and $i=6.34\cdot 10^1$ deg
the period of the perigee would amount to $\sim 10^7$ days, i.e.
$\sim 10^4$ years.
\section{Conclusions}
The Dvali-Gabadadze-Porrati braneworld model, in the Lue-Starkman
extension related to a spherically symmetric central mass, is very
interesting because it predicts, among other things, small
modifications of the Newton-Einstein gravity in the weak-field
approximation which have testable phenomenological implications
over the Solar-System lengthscale.

In this paper we have explicitly worked out its effects on the
Keplerian orbital elements of the orbit of a test particle without
restricting to circular orbits. It turns out that the pericentre
$\omega$ and the mean anomaly $\mathcal{M}$ undergo secular
precessions which are independent of the size and the shape of the
orbit. The sizes of these rates are $\sim \mp 4\cdot 10^{-4}$
\asec\ and $\pm 1.4\cdot 10^{-3}$ \asec, respectively. The first
nonvanishing corrections due to the eccentricity are of order
$\mathcal{O}(e^2)$. They are of the order of $10^{-5}$ \asec.

The possibility of observing such effects in the orbital motions
of the inner planets of the Solar-System has been examined, with
particular emphasis on Mars. The mean longitude $\lambda$ has been
considered along with various competing Newtonian and Einsteinian
effects. For $\lambda$ the braneworld shift amounts for all
planets to $\sim 1$ mas over almost one century while the
present-day observational accuracy, based on the processing of
almost 300000 data of various kinds for the EPM2004 ephemerides
spanning a 92-years temporal interval, is $3\cdot 10^{-3}$ mas for
Mars. A suitable linear combination with Venus would allow to
reduce the impact of the systematic errors.
The observational error would be $\sim 7\%$.

The possibility of measuring the LS pericentre precession in the
gravitational field of the Earth with the perigee of the proposed
LARES/WEBER-SAT satellite, although appealing, cannot be realized
because of the systematic errors due to the mismodelling in the
even zonal harmonics of the terrestrial gravitational field. Since
the LARES/WEBER-SAT is conceived to be a passive satellite,
without any active mechanism of compensation of the
non-gravitational perturbations, also such kind of systematic
errors, which the perigee of the geodetic satellites are
particularly sensitive to, would be fatal.
\section*{Acknowledgements}
I gratefully thank T.M. Eubanks for the useful material on the
asteroids  sent to me and for the helpful discussions on the role
of the orbital resonances induced by the asteroids on the Martian
orbit. Special thanks also to the anonymous referees whose
comments and observations were of great importance in improving
the manuscript.


\end{document}